\RequirePackage{amsmath}
\documentclass[runningheads]{llncs}
\usepackage{graphicx}

\usepackage{multirow}
\usepackage{color}
\usepackage{hyperref}

\usepackage{listings}
\usepackage{amsfonts}

\newcommand{\R}{{\mathbb R}}

\newcommand{\be}{\begin{equation}}
\newcommand{\ee}{\end{equation}}
\newcommand{\ba}{\begin{array}}
\newcommand{\ea}{\end{array}}
\newcommand{\baa}{\left[\begin{array}}
\newcommand{\eaa}{\end{array}\right]}

\newcommand{\beqa}{\begin{eqnarray}}
\newcommand{\eeqa}{\end{eqnarray}}
\newcommand{\bt}{\begin{tabular}}
\newcommand{\et}{\end{tabular}}

\newcommand{\bi}{\begin{itemize}}
\newcommand{\ei}{\end{itemize}}
\newcommand{\bc}{\begin{center}}
\newcommand{\ec}{\end{center}}

\newcommand{\code}[1]{{\footnotesize\texttt{#1}}}
\begin{document}

\title{Pinky: A Modern Malware-oriented Dynamic Information Retrieval Tool}

\author{Paul Irofti\inst{1,2}}


\institute{LOS-CS-FMI, University of Bucharest, Romania \and
Institute for Logic and Data Science, Romania\\
\email{paul@irofti.net}}

\maketitle
\begin{abstract}
We present here a reverse engineering tool that can be used
for information retrieval and anti-malware techniques.
Our main contribution is the design and implementation of
an instrumentation framework aimed at providing insight on the
emulation process.
Sample emulation is achieved via
translation of the binary code to an intermediate representation
followed by compilation and execution.
The design makes this a versatile tool that can be used for multiple task
such as information retrieval, reverse engineering, debugging,
and integration with anti-malware products.

\keywords{binary analysis  \and dynamic analysis \and information retrieval}
\end{abstract}
\section{Introduction}
In this paper we present the design and implementation of a new security
tool called Pinky.
Although an emulator at its core,
Pinky comes with its
own instrumentation framework,
intermediate representation,
coupled with a set of translators and compilers,
and platform emulation (filesystem, memory, libraries)
thus
allowing samples from multiple operating systems
to be analyzed and executed on any platform or machine.
For example,
its platform independence
allows it 
to analyze a Windows 32-bit executable on a Linux distribution
running on a MIPS-64 platform.

The instrumentation framework is designed such that
the emulation process can be stopped at any point
in order to provide data on the state it is in.
For example we can peak and change
mapped memory,
registers,
stack,
executable code and data sections,
and the filesystem.
At the same time,
we can also enable, set, or disable various information points.
The instrumentation framework has no performance impact on
the emulation process.
Through instrumentation,
Pinky can put on various hats.
It can act as a
tracer for system and library calls.
Suspending and resuming emulation
allows it to create memory dumps at various execution points,
making it an universal unpacker.
With more abstract instrumentation points,
the tool can also become a 
reverse engineering debugger.
And finally,
it can act as
anti-malware engine enabling the setup for
static and dynamic signatures through its callback mechanism.

Pinky is designed as an opaque tool
providing a clear and simple interface that
allows it to be integrated and controlled by third-party applications in
a non-intrusive way.
Thus it can be used inside datacenters (e.g. as an automated information
retrieval tool or scanner),
together with other software (e.g. existing IDS or anti-malware solutions),
or as a stand-alone reverse engineering tool inside a laboratory.

\noindent\textbf{Existing work.}
We focused our work on recent studies regarding dynamic binary analysis.
Our main inspiration for the intermediate representation and the
compiler-translator coupling has been the work on
UQBT~\cite{cifuentes2000uqbt,ung2006dynamic,troger2002analysis}
but also others like
\cite{nethercote2004dynamic,nethercote2007valgrind,henderson2014make}.

We differentiate ourselves from existing tools such as generic emulation tools
such as QEMU~\cite{bellard2005qemu} or Bochs~\cite{lawton1996bochs}
through performance, customization and instrumentation.
The goals are different, we do not plan on being a generic virtualisation
solution.
This is important,
especially in the anti-malware scenario,
because
speed difference and low-memory footprint 
sets us apart from generic solutions.
More sophisticated tools like Valgrind~\cite{nethercote2007valgrind}
or rr~\cite{callahan17_rr}
are more advanced in some regards,
but they do not offer platform customization, a file system,
instrumentation, nor cross platform emulation.

In the examples displayed in this paper we will see that,
through instrumentation,
Pinky can act as many well known tools.
For example,
it can give traces of system calls and native APIs
(such as ntdll.dll, kernel32.dll, advapi32.dll and so on)
much like \code{strace} and \code{ltrace} do in Linux.
But while it does that,
at the same time it can provide much more functionality.

\section{Emulator Schematics}

Following the work on intermediate representation languages
we seek to obtain a fast and performant emulator
through
our 
virtual machine (VM) implementation
coupled with just-in-time (JIT) compilation strategies
and efficient caching.
The main emulation performance is gained by tiered compilation
through threshold-based mixins of JIT compilation and VM emulation.
Every codeblock that gets processed is also cached
and will be reused the next time it's encountered.
Currently there are two caching strategies to choose from.
More aggressive optimizations can occur when a codeblock is frequently enough.
If a platform is missing JIT support, it will always fallback on the VM.

Instrumentation is done by dynamically enabling and disabling information
retrieval points throughout the emulation process.
Data points can adhere to dictated caller-callee protocols
and exchange data structures that influence
the sample control and data flows.
The instrumentation points have
no performance impact when they are disabled.

The emulation callback system is designed with the anti-malware engines in mind.
For example a common issue that comes up in the field is handling polymorphic
routines in static unpackers and coping with the different versions and
variations in the wild.
This can get to a point where the static routine gets so complex and has
to deal with so many cases that it slowly becomes a dynamic analysis tool on its
own.
In this sceanario,
a solution would be to
let the static unpacking process run until the offending polymorphic routine is
reached,
stop and handle things over to the emulator which will dynamically unpack it and
then give control back to the static routine.

When used as a reverse-engineering tool, it can act as a debugger by setting
breakpoints, watches, single-stepping at different granularity (e.g codeblocks
or instructions), setting different instrumentation points at runtime, tracing
system calls and library APIs, enabling different logs at various verbosity
levels for discrete periods of time, and many other similar useful features.
As a tool in the laboratory,
it can also be used in bulk scans to craft generic or specific reports.
The generated data can include
sample geometry,
memory dumps,
classification criteria,
profiling data 
and other custom data retrieved through instrumentation.

Reproducible results are made possible through the 
ability to stop the emulation process in a coherent and platform agnostic fashion.
This implies reproducibility no matter of the processor frequency,
memory size or type,
disk input-output throughput
or other machine dependent factors. 

Pinky is written in C++ with a focus on the C-subset with portability in mind.
The interface is simple and intuitive. It consists of three parts:
the emulator interface,
the configuration interface
and
the instrumentation interface.
It is implemented through abstract virtual classes that make it easy to
decouple from the rest of the project.
In the following sections we will describe each tool component and go into more details 
about its design and implementation.

\section{Design and Implementation}

\subsection{Intermediate Representation}

The goal of the intermediate representation (IR) was to to have a small and reusable
instruction set architecture (ISA) that would cater to all existing hardware and
software computer models.
In order to keep the instruction set small we designed an orthogonal ISA~\cite{Null14_orthisa},
thus allowing us to separate addressing modes and opcode
functionality~\cite{Jamil95_risc}.
Existing hardware examples are the PDP-11, VAX, and ARM11 architectures.
Orthogonality also allowed us to enforce fixed size instructions which in turn made it easier for us to enforce aligned access.
Our architecture address resolution is 32-bits,
and its instruction size is identical to its word size.
Each of our instructions has the following fixed form:
\code{opcode size dest src imm flags}.
This is unlike most hardware implementations~\cite{BGBF97_arch}
which have variable instruction size and permit unaligned access.
The x86 family~\cite{Intel19_SDM}
is particular famous in this regard as it permits constructs
like
\begin{lstlisting}
Before:                                                         After:
407F1A E834000000 CALL sample.407F53
...
407F4F 20978CEAF873 AND BYTE PTR DS:[EDI+73F8EA8C],DL           407F53 F8 CLC
407F55 020F ADD CL, BYTE PTR DS:[EDI]                           407F54 7302 JNB SHORT sample.407F67
\end{lstlisting}
The \code{call} instruction at address \code{1A}
jumps in the middle of the \code{and} instruction at address \code{4F}
which is interpreted as a legitimate but entirely different instruction.
Note that the entire program flow is affected by this and that
the attacker relied on the fact that a legitimate hidden instruction exists at
address \code{53}.

Stability was another key aspect because once we will start having
consumers of our architectures (also called translators),
it would become difficult to make large changes in the initial choice.
This is also true for compilers and interpreters that will use the resulting
intermediate representation to target code and execute it.
The ARM architecture is infamous for its frequent ISA changes.
A small and stable ISA meant that we had to ensure that we can reduce CISC
architectures to it. We did and we provide a few examples of difficult
instruction subsets (such as SSE and FPU) that we were successfully able to
emulate with our ISA in the following sections.
\begin{table}
\tabcolsep 5pt
\scriptsize
\caption{Instruction set}
\label{tab:xirins}
\centering
\bt[h]{rc}
Control & \code{jmp, ret, fsave, frestore}\\
Memory  &\code{ld, st, mv} \\
Arithmetic  & \code{add, addc, sub, subc, mul, div} \\
Logical  & \code{and, or, xor, not, cmp} \\
Shifts  & \code{rl, rr, sl, sr} \\
Special  & \code{syscall}
\et
\end{table}
With that in mind,
we are now able to define our cross intermediate representation (XIR).
In Table~\ref{tab:xirins} we present the entire instruction set.
The control instructions handle jumps, function returns, and flag manipulations,
while for memory manipulations we only have load, store and move instructions.
The arithmetic and logic operations consist of the usual suspects with the note
that the some have a \code{c}-suffix denoting an extra carry operation.
Shifts and rotation are supported also.

When designing such a tool,
if going after full CPU and thus ISA support
one of a few hard choices has to be made:
design only for a specific platform (e.g. IA-32-based only),
sprinkle hacks throughout the codebase thus ensuring multiple layer violations
(the translator reaches into the intermediate representation, or even directly
into the compiler),
or, in academic spirit, we can just ignore them and have a toy example working
only on an instruction subset.
In this article we propose an alternative approach which is able to deal with
all special architecture specific instruction set extensions.
That is why at the end of Table~\ref{tab:xirins} we introduced a special
instruction \code{syscall} that maintains modularity and solves the issue by
calling out to the emulator for help.
We have implemented and tested its usefulness with multiple extensions such as
Intel's FPU, MMX, SSE instructions.
We consider this to be fully extensible to others and also consider it future
proof.
Of course,
this instruction is slower.

When picking registers we went with 256 word-sized 32-bit registers
with 8-bit access.
Further,
we partitioned them into groups:
upper range mapped to the registers of the emulated architecture,
lower range reserved for compiler internal use,
plus other special registers for interrupts, flags and initialization.

In terms of stack choices,
while most architectures have a word-sized stack
or worse, a multiple granularity stack like x86,
we choose no stack at all.
This avoids multiple security issues
Even though the ISA has no concept of a stack nor does it emulate it in any way,
the stack of other models is modeled as direct memory access operations.
To our knowledge,
this represents a novel approach.

\subsection{Translator}

Each instruction set architecture that we want to support
has to provide
a disassembler
and
a translator to our intermediate representation.
The disassembler
tokenizes the instructions,
fetches the implicit or explicit opcode arguments,
and dispatches this information for translation.
Our translator interface consists of only two functions:
{\scriptsize \code{translate(mmu, addr, ir); syscall(env, mmu, opcode)}}.
The first translates block at address \code{addr} 
using the current  memory contents as reflected by the \code{mmu}.
and returns the intermediate representation \code{ir}.
For each opcode we have a translating function (or a handler) that
receives the opcode arguments
and
writes out the equivalent functionality in IR opcodes

A typical x86 opcode translation will look like
\code{gen\_opcode(dst, src, aux, mod)}
where the first three represent operands that can have
various types like register, memory, immediate value.
The last argument, \code{mod}, represents the instruction modifier that can
dictate a switch to a different addressing mode (e.g. 16-bit)
or a special request (e.g. repeating the instruction multiple times,
locking  etc.).
The function call
will generate a stream of equivalent XIR instructions.
\begin{lstlisting}
gen_add(dst, src, aux, mod)
  if (dst->type == OP_MEM && src->type == OP_IMM)
    reg_t tmp = alloc_reg();

    ld(dst->width, tmp, dst->r, dst->imm);
    add(dst->width, tmp, 0, src->imm);
    st(dst->width, dst->r, tmp, dst->imm);

    free_reg(tmp);
\end{lstlisting}
In the above we depict the x86 \code{ADD} translation where the destination
is the memory address of an integer to which we have to add an immediate value.
This translates to three XIR operations:
we load the integer value from memory to a temporary register (\code{ld}),
then we perform the addition (\code{add}),
and store back the result (\code{st}).
Notice that we used the destination width to dictate the addressing mode.
This makes the code portable and adaptable to word size changes.

As earlier discussed,
\code{syscall} provides instruction emulation for particular instruction subsets.
The registers and memory layout are prepared by the translator
before calling out to the compiler to solve the specifics of the 
\code{opcode} given the current execution environment.
Thus,
when encountering a special instruction the emulator will
pause and exit translation,
emulate (part of) the instruction on the real CPU,
write the results in translation state registers
re-enter and resume translation.
Here is a quick example for the x86 \code{FABS} instruction
\begin{lstlisting}
gen_fabs()                        emu_fabs()
  sys(UD_Ifabs, 0, 0);              double fpdata = FPU_ST(0);
                                    if (!isnan(fpdata))}
                                      FPU_ST(0) = fabs(fpdata);
\end{lstlisting}
Once the disassembler,
udis86 in this case~\cite{Vivek09_udis86},
decodes the instruction
it calls \code{gen\_fabs} from the translator in order to obtain IR.
This being a spceial FPU (or x87) instruction,
the event is marked through a \code{syscall} with the appropiate opcode id that
the complier will handle.
The IR is thus a single \code{syscall} instruction.
When the compiler reaches this instruction,
it ties it via the identifier to the special complier function \code{emu\_fabs}
that will know how to handle the special opcode
via x87 specific instructions as can be seen above.
Through similar \code{syscall} mechanisms,
the tool can also handle kernel (ring-0) sample emulation.

A special mention is required in regards to the handling of flags.
The flags register does not have a special status.
It is manipulated as any other register
and it is modeled in an architecture specific way by each translator.
Internal changes and checks can be protected by \code{fsave} and \code{frestore}
guards.
Post translation, the compilers are in charge of keeping the flags sound.
In particular,
the XIR virtual machine mimics the flag behavior of x86.

\subsection{Compiler}

Once everything is translated,
the XIR instructions can be executed via
interpretation,
compilation,
or a mixture of the two (also called tiered compilation).
Interpretation is done through the XIR virtual machine (XIRVM).
The implementation is straight forward:
for each IR function (see \code{ADD} example translated above)
we execute each XIR instruction in the emulator's own process space.
The sample is isolated in a memory mapped region where all XIRVM operations
perform their tasks.
Note that we only need to implement a few VM instructions; the ones listed in
Table~\ref{tab:xirins}.
\begin{lstlisting}
exec_st(mmu, env, pc, dst, src, imm, flags)
  b =  flags & BITS_MASK;
  addr = env->regs[dst] + imm;
  val = read_reg(env, b, src);
  size = 1 << b;

  set_word_le(&val, val);

  page = mmu->pte[addr >> PAGE_BITS]
  offset = addr & (PAGESZ - 1);
  if (page != 0 && offset + size <= PAGESZ)
    memcpy((uint8_t *) page + offset, &val, size);
  else
    mmu->write_memory(addr, &val, size))
\end{lstlisting}
In the above example we depict the XIR store (\code{st}) virtual machine interpretation.
The first instructions fetch the addressing mode in \code{b},
the memory address from the destination operand,
and the value to be written from the source operand.
Based on the size \code{b} we store the read value in proper endianess and
alignment according to the target architecture.
Here we assume it is word sized, but it can be any subdivision or multiple of
it.
Next,
the memory address is translated into a page and offset within the virtual
machine memory management unit.
If the page is already mapped, we perform a simple memory copy instruction.
Otherwise we call out to the MMU to perform the write,
which also implies a page mapping operation beforehand.

Compilation is performed via just-in-time (JIT) compilation strategies.
Similar to the interpreter,
each IR function is compiled and executed natively on the host machine.
Implementation is also straight forward due to the reduced number of
instructions in the XIR ISA.
We tested with several JITs for both 32-bit and 64-bit targets.
For x86 we used AsmJIT~\cite{Kobalicek11_asmjit}.
\begin{lstlisting}
gen_mv(dst, src, imm, flags)
  b = flags & BITS_MASK;

  switch (b)
  case B32:
    if (src)
      as.mov(eax, XIR_REG32(src));
      if (imm)
        as.lea(eax, dword_ptr(eax, imm));
      as.mov(XIR_REG32(dst), eax);
    else
      as.mov(XIR_REG32(dst), imm);
    break;
\end{lstlisting}
In the above example we depict the XIR move (\code{mv}) JIT compilation.
The first instructions fetch the addressing mode in \code{b}
and in the displayed operations we assume it is 32-bit,
but it can obviously be any other mode.
If the source operand is defined,
we have to emit a register-register move instruction.
If only the immediate value \code{imm} is defined,
then we move it to the destination register and we are done.
If both the source and the immediate operands are defined,
then we treat it \code{imm} as an offset from \code{src}.

As with other systems,
the interpreter is generally slower than the compiler.
But often we found that when a XIR function is not repeatedly called,
the effort of compiling the code outruns the gain in running native code.
Thus in these cases it might be better to just use the interpreter.
To handle this scenario we implemented tiered compilation~\cite{Bebenita10_tiered,Hartmann14_tiered},
where the IR is compiled only if its usage passed a certain threshold. 
In order to improve the performance of the translate-compile cycle,
we added caching for IR functions
such that already codeblocks that have already been processed can go
straight to execution.

\subsection{Memory Management Unit}

Earlier we saw memory store operations,
What happens when any of the following needs to be emulated:
{\scriptsize\code{MOV EAX, [1000]; JMP [EDX]; STOS DWORD PTR ES:[EDI]}}.
The instructions alone can not describe the entire system state,
we need to keep track of memory writes and reads.
This involves having 
an initial memory state before starting the emulation process.
This initial state is operating system (OS) dependent.
The stack state is also partially dictated by the OS in general,
and by its C library implementation and by its format for executables.
Thus
doing writes and reads forces us to set and maintain an internally stored memory
map.

To address these issues,
we designed a transparent platform-agnostic memory management unit (MMU).
Its contents is data without any semantics or logic tied to it.
We choose to represent it as a flat 4096-bytes paging system
such that memory access can be done with $O(1)$ complexity.
The memory is allocated contiguously and grouped into memory regions.
These are automatically managed by the MMU when memory is allocated or freed by
the emulator.
A caching mechanism is set in place in order to take a big load
off of the translator and the compiler resulting in big speed-ups.
Overall this makes it a performant and clean memory representation.
\begin{lstlisting}
read_memory(va, buffer, size);            pmap(sz, perf_va, actual_va, flags, min_va, max_va);
write_memory(va, buffer, size);           pmap_lookup(count, pref_va, min_va, max_va);
void dump(dmp_dir);                       pmap_remove(start_va, end_va);
\end{lstlisting}
The interface is simple and similar to what system programmers are used to
encounter when dealing with memory.
The first functions map pages into memory;
\code{pmap} wires the required pages for a \code{sz} sized buffer
with optional constraints such as virtual address (\code{va}) interval
or forcing a fixed mapping via \code{pref\_va} and \code{flags}.
Calls to the read and write memory operations were presented earlier in the 
compiler section;
the functions require a virtual address, the buffer and its size.
Finally, 
\code{dump} is a very useful function to be called at various emulation points
in order to inspect the memory layout and its contents.
It can be used for malware analysis, information retireval or debugging tasks.

\subsection{File System}

With an MMU,
we still have to address other memory problems during execution.
Consider the following sequence that can appear in our emulated sample
\begin{lstlisting}
01002E8D PUSH ESI
01002E8E LEA EAX, [EBP-0x8]
01002E91 PUSH EAX
01002E92 CALL DWORD [0x1001074]
7DD85AB0 CALL DWORD 0x7dd85ab5
\end{lstlisting}
representing an API call to \code{kernel32.dll!GetSystemTimeAsFileTime};
a function implementation inside a shared system library.
These are usually stored as imports
inside a special section of the sample's executable
in respect to the executable format of the underlying operating system.
Almost all executables have at least a few such imports in order to function
properly.

The same issue arises when the sample wants to access the file system for common
input-output (IO) operations such as
creating, reading, or writing to files and directories.
In Windows operating systems it might even call out to manipulate registry
entries,
or similarly on Linux touch and modify \code{/proc} entries.
While we can emulate or get around some of these issues,
most calls do not have a clean solution and thus require the presence of a file
system.

We address this issue by creating a virtual file system (VFS)
that
stores created or modified files throughout the emulation process.
In addition it provides a minimal file system environment resembling the
expected OS
and it also takes care of special features such as registry
and
mimics special files such as the ones found in \code{/proc} and \code{/dev}.
Thus
VFS provides an interface for creating and managing
file system containers that are platform specific and that are generated before
the emulation process through an archiving like tool.
\begin{lstlisting}
init(container);                         unlink(path);
fd = open(path, mode);                   stat(path, size, attributes, mode, base);
close(fd);                               seek(fd, pos);
read(fd, buffer, size);                  chmod(path, attributes);
write(fd, buffer, size );                rename(from, to);
\end{lstlisting}
After loading the container with \code{init},
the VFS interface follows the UNIX system call conventions for handling files.

\subsection{Executable Loader}

With the system memory and file system present,
the final missing puzzle is the executable loader.
Without it the API call problem still exists:
a connection between the sample and the library needs to be made
and that link is present in the sample file.
Each executable follows an executable format depending on the operating system.
The executable format dictates how the file is partitioned into sections.
The sections contain information about external dependencies,
including libraries and the functions therein used by the current sample.
Thus,
a loader should
setup the virtual address space, including the stack, for the sample
and
resolve links to external libraries.

For popular platforms such as Linux, BSD or Mac
that use the ELF format~\cite{Lu95_ELF,Shapiro13_WeirdELF},
open-source implementations exist that can be integrated in the emulator.
Windows uses a similar but different format called portable executable (PE)~\cite{Pietrek94_PE}
that is mostly undocumented and depends on the kernel version.
Given the wide impact of malware and other malicious software on the Windows platform,
we also designed and implemented a PE loader.
Our PE loader
mimics as close as possible the NT kernel,
passes all non-conforming but loading samples we found in the wild,
and passes all tests on the Corkami dataset~\cite{Albertini_PEdataset}.

When providing the actual library implementations,
existing solutions either
emulate the real functions
and run them outside emulation
or
use external binaries, perhaps the exact platform library binaries,
and run them inside the emulator.
Because of the delicate subject of distributing external binaries,
but also the man-hour impact of rewriting the existing ones,
we chose to provide both options.
The emulator will try the native implementation and,
if it can not find the function,
it will try to find the binary in the VFS and load it.
Of course,
for internal laboratory use it is enough to create a file system container
with the original libraries
which is completely possible via the VFS functionality.

Writing your own native implementations does come with advantages
such as
the fact that you trust the code (since you wrote it)
and can thus gain extra performance by running it outside emulation.
Also, in general, the implementations are simpler and smaller in size.
The down sides are the fact that
running it outside of emulation means that
if it crashes it brings the entire process to a halt
and it is also harder to debug.

Using external libraries wrapped in a VFS container has the
advantage of having each library call going through the emulation 
layer and thus gaining better control and insight on the whole process.
Also,
crashing does not affect the emulator.
The down sides are increased complexity due to emulation and running
through abstractions that might not be needed for the task at hand.
A windows library has to account for many use-cases and inter-connections
and comes with no redistribution rights.

\subsection{Instrumentation and Information Retrieval}

Dtrace is a modern dynamic tracing tool~\cite{Mcdougall06_dtrace}
used in most modern operating systems~\cite{Gregg11_dtrace}
for debugging, accounting, logging and other information retrieval tasks
such as reverse engineering~\cite{Beauchamp08_REdtrace}. 
Unlike most tools,
dtrace has the advantage of having zero cost when disabled,
a feat accomplished through machine dependent tricks.
This allowed for the spread and setup of
multiple instrumentation points (or probes) at no cost.
When needed,
these information points can be enabled and executed (or fired up).

In our emulator we followed the dtrace model and implemented a similar
functionality across all modules.
The probes have no cost when setup and can be fired at any time during
emulation.
Once implemented,
this enabled us to quickly gain useful features such as
feedback at any point during emulation,
peaking at
mapped memory,
registers,
stack,
executable sections,
and the file system.
The instrumentation framework has no performance impact on the emulation.

\begin{lstlisting}
probe_enable(probe_id);         probe_create(probe_id, name, provider, enabler);
probe_disable(probe_id);        probe_register(probe_id, consumer, consumer_id);
                                probe_cb_consumers(probe_id, context);
\end{lstlisting}
We defined the probe interface is as follows.
A probe has a provider and multiple consumers.
Once a provider creates a probe,
a consumer can register
using the probe unique identification number or the probe name.
Registered consumers are walked through when a probe is fired either from
the probed function itself or through a generic consumers callback.
If the probe has a broadcast-like functionality,
the later is preferred.
If a certain list of conditions need to be fulfilled for a consumer trigger to
be pulled,
then the former is the way to go.

\section{Results}

\subsubsection{Information retrieval}
Through the use of the instrumentation probes,
we built a flexible configuration framework
that, during emulation, allows us to
change (with immediate effect)
all the emulation options,
tweak the interpreter, compiler and the tiered compilation threshold,
and also switch caching algorithms.
Through the same configuration interface
we support multiple level logging for all of the emulator's modules
that can can be turned on, off or switched to a different verbosity at any time.
\begin{lstlisting}
010029E3 push ebx  ----------------+  ST32 [r165-0x4], r164
010029E3                           -  MV32 r165, r165-0x4
010029E4 push edi  ----------------+  ST32 [r165-0x4], r168
010029E4                           -  MV32 r165, r165-0x4
010029E5 call dword [0x1001058]  --+  MV32 r165, r165-0x4
010029E5                           +  MV32 r32, 0x10029EB
010029E5                           +  ST32 [r165], r32
010029E5                           +  LD32 r32, [0x01001058]
010029E5                           -  RET r32 
\end{lstlisting}
In this example we turned on IR debugging to see how the translator
turned x86 machine code (left hand side)
into XIR instructions (right hand side).

We also have a probe interface that can
stop the emulation process
and feed memory regions through the MMU
to static analysis tools for further insight.
Based on these results, 
the external tools can change the behaviour or control flow of the analyzed
sample
before resuming emulation.

\subsubsection{Command line debugger}
We put together multiple probes to
create a command line tool for inspecting, controlling and changing the
emulation process.
This tool includes debugger functionality like setting breakpoints,
watchpoints and more sophisticated conditional stopping points
all through the use of probes.
This tool can also produce on-demand MMU dumps during emulation for signature
inspection.
\begin{lstlisting}
> break 0x7DE9FA40
> ping.exe
EMULATING ping.exe
Breakpoint 0 at 0x7DE9FA40
> probe x86_step_mode
> set log:ir 1
> next
DEBUG - debug_code.cpp:301 - ir:
Source -> IR:
7DE9FA90 mov dword [ebp-0x10], 0xffffffff  +  MV32 r32, 0xFFFFFFFF
7DE9FA90                           +  ST32 [r166-0x10], r32
7DE9FA90                           -  RET 0x7DE9FA97 
Breakpoint 1 at 0x7DE9FA90
\end{lstlisting}
Above is an example inside the debugger.
First we set breakpoint at an address inside the Windows \code{ping}
executable and then proceed to run the sample.
The debugger stops when the address is reached.
Then we set fire the stepping mode probe that turns every codeblock into
a single instruction,
enable the logging level for the IR translation
and proceed to the next instruction.

\subsubsection{Stopping}
We provide deterministic stopping that is agnostic to the host hardware.
The goal is to be able to stop the emulation process
aronud the same instruction no matter if we run on
an Intel Xeon or a small ARM device.
To do that we started an ample analysis where
we marked the important nodes in the dynamic analyzer,
added counters in these key positions,
ran the emulator through large corpus of varied data samples
and at the end
stored the execution time and the final counter values.
The corpus consisted of $m$ samples
with $n$ counters each
such that $m \gg n$.
Thus a given sample has an execution time
$
t = \left( \ba{ccccc} c_1 & c_2 & c_3 & \dots & c_n \ea \right)
\left( \ba{ccccc} w_1 & w_2 & w_3 & \dots &  w_n \ea \right)^T
$.
Let $T\in\R^{m}$ be the vector consisting of all sample execution times,
$C\in\R^{m\times n}$ the counters matrix and $w\in\R^{n}$ the weights.
These measurements lead to a simple least-squares problem~\cite{Golub12_matrix}
$T=Cw$
whose solution are the associated weights $w$.

This model leads to some nice practical properties.
We can start with a small set of counters which leads us to a fair approximation
thus gaining a fast start-up.
This starting point can be continuously adjusted and improved
through counter addition and elimination
but also through the addition of new sample information.
This can also be seen as a profiling tool.

We name the weight values metrics.
The speed of a platform is measured as metrics per second.
We can now build a deterministic threshold
by
computing only once an average platform speed,
and setting a metric threshold based on that.
If a process was stopped we know exactly where.
As a side effect,
we also get an implicit time threshold for free.
For example,
if we have an average platform speed of 50 metrics per second,
we can set the threshold to 150 metrics
which results in a 3 second maximum emulation time per sample.

\subsubsection{Production}
The tool has been integrated and used successfully
in an anti-malware engine environment
(acting as a generic unpacker and memory inspection
tool and doubling the product detection rate),
as a bulk scanning tool for malware and clean sets,
and also as a debugger-like reverse-engineering
tool for sample analysis.
Three applications that seamlessly integrated the library with success.
This lead to a few nice properties, software wise.
The emulator
is reentrent
and has built-in exception and fault protection
for POSIX and Windows operating systems.
Through continuous integration,
it is tested weekly on $1,000,000+$ samples
with support for multiple debugging and quality assurance tools
such as OProfile and Electric Fence.

The emulator is highly portable.
For example the bulk scanning tool runs on Linux, OpenBSD and Windows
with 32-bit and 64-bit Intel-derivate CPUs.
Also quick nightly scans are conducted on a wide range  of system configurations,
both big endian and little endian, with hardware platforms such as Intel 32-bit
and 64-bit, ARMv5 and ARMv7, MIPS-64, PowerPC, Sparc, Sparc64, HP-PA,
and on operating systems such as Windows (versions from Windows XP up to
Windows 10), OS X, Linux, FreeBSD, OpenBSD, NetBSD, Solaris, IllumOS, Darwin and
others.
The solution is compiled with all mainline compilers:
Visual Studio, GCC, and CLang.

\section{Conclusion and future work}

In this paper we presented a reverse engineering tool that can be used
for information retrieval and anti-malware techniques.
Our main contribution has been the design and implementation of
an instrumentation framework created to provide insight on the
emulation process that is achieved via
the translation to an intermediate representation
and then compilation of the studied sample.
In the results section we show-cased its application to multiple tasks
such as information retrieval tool, debugger and its ability to integrate in an
anti-malware production environment.
Due to the reduced number of instructions in the XIR ISA,
adding translators and  JITs is not a difficult task
which makes us consider adding an LLVM translator
in the near future.

\bibliographystyle{splncs04}
\bibliography{bib}

\end{document}